# An Inhomogeneous Josephson Phase Near the (Super) Conductor-Insulator Transition

Jiufeng Tu[*] and M. Strongin, Brookhaven National Lab, Upton, NY, 11973

Y. Imry, CM Physics, the Weizmann Institute of Science, Rehovot, Israel, 76100

## Abstract

*In many cases inhomogeneities are known to exist near the metal (or superconductor)- insulator transition, as follows from well-known domain-wall arguments. If the conducting regions are large enough, and if they have superconducting correlations, it becomes energetically favorable for the system to go into a Josephson- coupled zero-resistance state before (i.e. at higher resistance than) the material becomes a "real" metal. We show that this is plausible by a simple comparison of the relevant coupling constants. We also illustrate using data in the literature on oxide materials as well as ultra-thin films, that when this proposed "Josephson state" is quenched by a magnetic field, an insulating, rather then a metallic, state indeed appears.*

Pacs. Numbers: 74.40.+k, 74.50.+r, 74.80.-g

Since the discovery of high-temperature superconductors there has always been the intriguing connection between underdoped high $T_c$'s and the properties of disordered and granular superconductors [1,2]. We argue in this letter that near the superconductor-insulator (S-I) transition, inhomogeneities lead to a zero resistance Josephson coupled state, which exists both in high temperature superconductors and "usual" superconductors even though the interactions [3] causing the superconducting state may indeed be very different. The underlying principle is that disorder implies inhomogeneities on *some* length-scales, as was first argued, in this context, by Kowal and Ovadyahu [4]. These scales depend on the nature and strength of the disorder. This picture is supported by numerous experiments [4,5] and may be related to domain formation by random-field-type impurities [6]. For example, if the Mott-type metal-insulator transition were in fact first order, as originally argued by Mott, then the arguments of Ref. 6 would imply "domain" formation in effectively 2D systems even for the weakest strength of the impurities! Finite-strength impurities will generically lead to domain formation in most situations, except very close to an appropriate second-order transition where the correlation length diverges strongly enough [7]. Experiments considering the effect of inhomogeneities brought about by fluctuations in the local electron density or concentration gradients already exist in the literature [5]. On the theoretical side, the importance of inhomogeneities has been highlighted by Emery, Kivelson and co-workers [8], and by Dagotto and co-workers [9]. Ghosal, et al. [10] have considered a model based on the Bogoliubov-de Gennes equations, of how "homogeneous" disorder introduces an inhomogeneous pairing amplitude in ultra-thin films. We would like to add to these interesting models that in the non-superconducting state, the phases of these domains are not locked and therefore the phase fluctuations should average the local pairing amplitude, $\Delta$, to zero (however, $<|\Delta|^2> \neq 0$). Refs 11 and 12 discuss Bose-Hubbard models and cite earlier theoretical references related to disordered systems. Other theoretical approaches will be mentioned later in the paper.

In recent work on ultra thin films [13] it was argued that in an inhomogeneous medium it is possible for a Josephson coupled superconducting state to be more stable at or near the S-I transition boundary (more disorder/less carrier density) than the metallic state (which is defined here as being on the metallic side of the percolation transition). This general problem was treated some years ago in Ref.14 using considerations based on the Thouless [15] arguments for the onset of localization in 1-D and handling the Coulomb effects in the spirit of the phenomenological arguments of Abeles and Shen [16] and Kawabata [17]. A simple case where this clearly works is an array of Josephson coupled clusters with an energy gap $\Delta$ that is larger [14,18] than the energy level spacing in the cluster (see below). In this present note we are interested in extending these ideas to give some insight into weak superconductivity in inhomogeneous systems, and thus whether we can understand data in films as well as in underdoped high $T_c$ superconductors.

Before dealing with the experimental data we briefly describe the simple argument which indicates that in an inhomogeneous system there is a regime in which a Josephson coupled state occurs before the metallic state, as the sample resistance decreases from a resistance characteristic of the insulating state to that of a normal metal. This is done by either increasing the doping in the high $T_c$ case, or changing the thickness in the ultra thin film case.

The Thouless picture of localization in one dimension can be generalized to analyze the electronic couplings between "metallic regions" [15] in an inhomogeneous system (which can consist of grains or doped regions with high conductivity) in any dimension [19,20]. The intergrain coupling energy is given by $\hbar/\tau$ = $V_L$ where $\tau$ is the lifetime for an electron in one of the conducting regions, of size L, to go into the next one. The conductance between "grains" can be related to the ratio of this coupling energy to the energy level spacing in the grains and is written as a dimensionless conductance, $g = V_L/w_L = 2\hbar/e^2 R_L$, where $w_L$ is the characteristic energy level spacing in the small metallic regions. When the typical intergrain resistance, $R_L > 2\hbar/e^2$ then the noninteracting system becomes localized. It is interesting that if one uses the analysis of Abeles and Shen [16] to estimate the resistance between isolated grains where the coupling energy overcomes the intergranular Coulomb energy of $e^2/2C$, then by approximating $\hbar/\tau$ as $\hbar/RC$, and setting this equal to $e^2/2C$, one gets the same value of $R \cong 2\hbar/e^2$ for the resistance below which the "intergranular" coupling is greater than the Coulomb repulsion (where R is the tunneling resistance between grains and C is the capacity of the two grains where d is the grain size). Thus in this case a system with Coulomb interactions will also be metallic once $R_L < 2\hbar/e^2$. Clearly, these two approaches are not unrelated. A physical argument relating them might be based on the fact that once the single-electron eigenfunctions are delocalized and spill over from the grain, the Coulomb blockade picture with quantized charge on the grain becomes meaningless. Evidently, this argument is certainly valid when the Coulomb energy is weak, $e^2/2C < w_L$, and it is treated as a perturbation on the noninteracting picture. The argument may also hold for strong interactions, $e^2/2C > w_L$, provided that the actual value for $R_L$, which may be strongly renormalized by the interaction, is used.

In Ref. 14 the noninteracting picture was generalized to include the effect of *strong* Coulomb interactions (i.e. $e^2/2C > w_L$), which of course is typically crucial due to the marginal screening and the charging energy when electrons move between conducting regions. Here, to get metallic behavior, the intergrain transfer energy should overweigh the Coulomb energy [14,16,17,19,20]

$$zV_L = 2\hbar w_L z/e^2 R_L > e^2/2C, \qquad (1)$$

where z is a coordination number, related to the typical number of nearest neighbors, which appears in the mean-field theory for the transition. The condition for superconductivity is however that the Josephson energy, given by the standard expression $E_J = \pi\hbar\Delta(0)/4e^2 R_L$, be larger than the Coulomb energy, or, putting again the factor z for a medium composed of grains we replace $E_J$ by $z E_J \sim zV_L\Delta(0)/w_L$.

$$z\pi\hbar\Delta(0)/4e^2 R_L > e^2/2C. \qquad (2)$$

Here $\Delta(0)$ is the gap at T=0. In other words, the pair transfer matrix element $E_J$ replaces here the single-electron coupling energy $V_L$. For this approximate argument at low temperatures, there is no need to put in the temperature dependence of the gap.

The interesting result is that there clearly exists an unusual regime where $E_{coul}$ can be greater than $zV_L$, but less than $E_J$, as long as $\Delta(0)/w_L$ is greater than 1. So for grains that are "large" in the sense [18] that $\Delta(0)/w_L \gg 1$, *superconductivity is easier to achieve than normal* conductivity [21]. This argument is only meant to show that if there are intrinsic inhomogeneities and the system has superconducting regions, then it is possible to have a Josephson state *before* having a metallic one.

A possible phase diagram, for $\Delta(0) \gg w_L$, is described in Fig.1. It can be seen that at low temperatures as the conductivity increases, (by increasing the thickness in the case of films and increasing the doping in underdoped high $T_c$'s) one first goes from the insulating phase into the Josephson phase (line A-B) and finally into the true metallic/superconducting phase where not only has percolation occurred, but the respective "wavefunctions" became delocalized [22]. (Note that line C-B and its continuation to higher temperatures, eventually becomes a smooth crossover rather then a sharp transition. We do not know whether the change from the Josephson phase to the percolating superconductor is effected by a real transition). This is consistent with the above argument showing that if there are superconducting correlations in the insulating regime, then the quantum transition to a Josephson state can occur (for large "grains" where $\Delta(0)/w_L >1$ ) *before* the percolation-delocalization transition. A crucial point, of course, is whether experimental evidence exists for our conjecture. Below we discuss the evidence for this in both ultra thin films and underdoped high temperature superconductors.

In the case of ultra thin films, Ekinci and Valles [23] and Hsu, et al. [24] have done two kinds of experiments that are relevant to this question. First they have performed STM measurements which provide evidence that their ultra-thin films were not a uniform amorphous material and the morphology changed as a function of film thickness [23]. They have also found [24] that in PbBi films where $R_\square$ is of the order of $h/4e^2$ and $T_c$ is depressed to near 1K, the application of a magnetic field on the order of a few Tesla suppresses superconductivity and $dR/dT < 0$ with a temperature dependent resistance change that is significantly larger than that expected from weak localization theory, and the behavior therefore suggests an insulating state. Both of the above observations are consistent with intrinsic inhomogeneities. For example, in zero field there can be a "Josephson" state, but in a magnetic field the weak Josephson currents, which link the superconducting regions are destroyed. This leads to disconnected superconducting regions in a normal matrix and an accompanying activated conduction, where $dR/dT <0$. We will come back to this point in somewhat more detail when the field dependence of underdoped high $T_c$'s will be discussed.      This possibility of a Josephson Phase in films is consistent with recent measurements of the optical conductivity in ultra thin films [13]. In this case it was found that in thin Pb and Au films deposited onto a- Ge at 10K, an optical anomaly occurred at 2000 –3000Ω/□. This is in contrast to the S-I transition in Pb films which occurs near 6500 Ω/□ or $h/4e^2$, (a value given by dirty boson theories) [12]. It was suggested that this difference occurs because at 10K (above the superconducting transition for Pb) the optical anomaly is a result of the actual percolation transition to the metallic state. At this transition the dielectric constant has a singularity and this is what causes the optical anomaly [25]. Of course, this measurement must be made above the superconducting transition where the insulator crosses over to the metal (i.e. above B in the extension of line C-B)

There are many phenomena in high temperature superconductors and other oxide materials that can be interpreted by models of inhomogeneities [8,9]. For example, recent STM measurements by Lang, et al. [5] have established that Bi-2212 behaves as a "granular" superconductor with domains on the order of 30A. As another possible example of inhomogeneities, there is a series of important papers by Ando, et al. [26] who measured the field dependence of the transition temperature and the normal state resistance of LSCO. In the underdoped samples, and only the underdoped samples, the interesting result is that superconductivity is suppressed in their field of about 61T and a lnT dependence of the resistivity follows. At higher temperatures (in the normal state) the magnetic field does not change the linear dependence of the resistivity except close to $T_c$. The lnT dependence is not fully understood and whether it is a property of a percolating inhomogeneous system or whether it is due to a new ground state in a magnetic field, which might have charge ordering or stripes, remains unknown. Recently Beloborodov, et al. [27] have shown that this lnT dependence occurs in granular systems and they mention that this could explain the above result on high temperature superconductors, if the samples are indeed granular or inhomogeneous. This is consistent with the lnT behavior **is** found in specially prepared inhomogeneous systems. For example, this lnT dependence has also been found in granular cermets of NbN in a boron nitride matrix [28]. When a large magnetic field is applied to destroy superconductivity in this system, a lnT behavior of the resistivity is found, very much like the work discussed above.

It is also interesting that Shahar and Ovadyahu [29] have observed a "phase diagram" similar to Fig. 1 in studies of indium oxide films. In this work they depict both insulating and superconducting behavior as a function of $k_f\ell$. What is particularly interesting is that the insulating phase boundary intersects the superconducting phase in a way similar to fig.1 and there is a region at small $k_f\ell$ that is identified as a coexistence of insulating and superconducting regions. This region was larger than predicted for a uniform system and the effect of inhomogeneities was discussed. In our interpretation here we would call this the Josephson phase.

In summary we propose the above picture of spontaneously formed conducting domains which form a Josephson phase at low temperatures, as a general property of disordered systems near the superconductor-insulator transition, and especially in the effectively 2D case [6]**.** We believe there should often exist intrinsic inhomogeneities near the S-I transition and have discussed various examples in the literature. These inhomogeneities dominate the initial transition to the superconducting state. As far as we know, there is really no experimental evidence for a uniform state at the S-I transition. In this regime where superconducting regions first appear (to the left of line A-B) it is plausible that they are initially decoupled (this region which is analogous to the pseudo-gap state is not shown in Fig.1). As line A-B is approached, Josephson coupling produces phase alignment of the order parameter of different superconducting regions, and this happens before the percolation-delocalization transition to the metallic state, along line C-B. The possible relevance of the inhomogeneous phase formed in the normal metal-insulator crossover regime to the "bad metal" behavior in high $T_c$ materials is an interesting question for further study.

## Acknowledgements

We thank Amnon Aharony, Joe Bhaseen, John Chalker, Sasa Dordevic, Alexander Finkelstein, Chris Homes, S.A. Kivelson, Robert Konik, Zvi Ovadyahu, Alexei Tsvelik, John Tranquada, Tonica Valla, Jim Valles, Matthias Vojta, and Peter Wölfle, and for helpful discussions.

Work at Brookhaven was supported by the Department of Energy (DOE) under Contract No. DE – AC02-98CH10866. Y. I. was supported by a Center of Excellence of the Israel Science Foundation (ISF), and by the German Federal Ministry of Education and Research (BMBF) within the framework of the German -Israeli project cooperation (DIP).

# References

*Present address: Department of Physics, The City University of New York, New York, NY 10031

**Figure Caption**

The figure indicates the "phase diagram" of an inhomogeneous superconductor in the Temperature/ Conductivity plane. The conductivity is used as a measure of disorder and increases with doping or film thickness. The A-B line is the boundary between the insulating phase and the Josephson coupled state where there are isolated superconducting regions that are Josephson coupled. The B-C line is where the system goes into a bulk superconducting phase (there are percolation paths). In this region to the right of line B-C we would expect normal metallic conduction when superconductivity is quenched by a magnetic field. In the Josephson phase a logT behavior in the resistivity seems common when superconductivity is quenched by a field. The note in the figure indicates a region to the left of line A-B where there exist disconnected metallic regions. Likewise, disconnected insulating regions occur to the right of the line A-B.

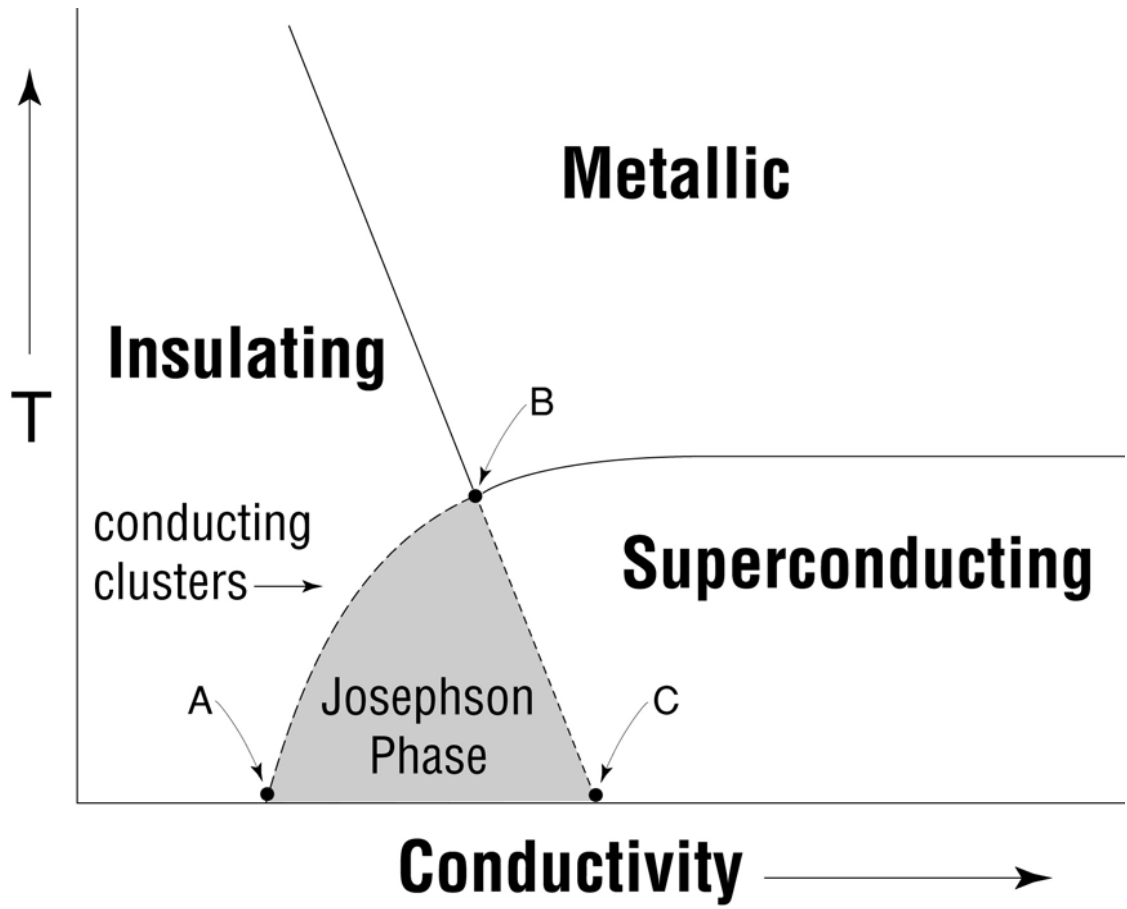